\documentclass{iopart}
\usepackage{epsfig}
\usepackage{subfigure}
\usepackage{ulem}   %I'm told this enables strikethroughs, handy for editing- SAW
\usepackage{cite}

\begin{document}
%%%%%%%%%%%%%%%%%%%%%%%%%%%%%%%%%%%%%%%%%%%%%%%%%%%%%%%%%%%%%%%%%%%%%%%%
% comment out for single column
%\twocolumn[%
%\hsize\textwidth\columnwidth\hsize\csname@twocolumnfalse\endcsname
%\draft
%%%%%%%%%%%%%%%%%%%%%%%%%%%%%%%%%%%%%%%%%%%%%%%%%%%%%%%%%%%%%%%%%%%%%%%%
\newcommand{\figwidth}{0.95\columnwidth}
\newcommand{\ffigwidth}{0.4\columnwidth}

\title[Comparative analysis of rigidity across protein families]{Comparative analysis of rigidity across protein families}

\author{S A Wells$^{1}$, J E Jimenez-Roldan$^{1,2}$ and R A R\"{o}mer$^{1}$}
\address{$^1$ Department of Physics
  and Centre for Scientific Computing, University of Warwick, Coventry,
  CV4 7AL, United Kingdom}
\address{$^2$ Department of Systems Biology, University of Warwick, Coventry,
  CV4 7AL, United Kingdom
}\ead{r.roemer@warwick.ac.uk}

\date{$Revision: 1.3 $, compiled \today}
%\date{\today}

\begin{abstract}
$Revision: 1.3 $, compiled \today\\

% Rigidity analysis using the ``pebble game'' has been applied to protein crystal structures to obtain information on protein folding, assembly and the structure-function relationship. However, previous work using this technique has not made clear how the set of hydrogen-bond constraints included in the rigidity analysis should be chosen, nor how sensitive the results of rigidity analysis are to small structural variations.
We present a comparative study in which ``pebble game'' rigidity analysis is applied to multiple protein crystal structures, for each of six different protein families.
We find that the mainchain rigidity of a protein structure at a given hydrogen-bond energy cutoff is quite sensitive to small structural variations, and conclude that the hydrogen bond constraints in rigidity analysis should be chosen so as to form and test specific hypotheses about the rigidity of a particular protein.
Our comparative approach highlights two different characteristic patterns (``sudden'' or ``gradual'') for protein rigidity loss as constraints are removed, in line with recent results on the rigidity transitions of glassy networks.
\end{abstract}


\submitto{\PB}
\maketitle
%%%%%%%%%%%%%%%%%%%%%%%%%%%%%%%%%%%%%%%%%%%%%%%%%%%%%%%%%%%%%%%%%%%%%%%%%%
%] % comment out for single column
%%%%%%%%%%%%%%%%%%%%%%%%%%%%%%%%%%%%%%%%%%%%%%%%%%%%%%%%%%%%%%%%%%%%%%%%%%
%\narrowtext
%\tighten

%%%%%%%%%%%%%%%%%%%%%%%%%%%%%%%%%%%%%%%%%%%%%%%%%%%%%%%%%%%%%%%%%%%%%%%%
\section{Introduction}
\label{sec-intro}
%%%%%%%%%%%%%%%%%%%%%%%%%%%%%%%%%%%%%%%%%%%%%%%%%%%%%%%%%%%%%%%%%%%%%%%%

It is a common goal in biophysics to represent the flexibility of a protein and study its large-scale motion without incurring the full computational cost of molecular dynamics simulations. One popular family of approaches is based on normal-mode analysis applied to a full or simplified representation of the protein structure
%from GohKC04:
\cite{CanSM02,%28
Cas99,%40
BahAE97,%41
MinKWM03,%42
MinKWBM02,%43
TamWB02,%44
Hal02,%45
%Elnemo
Tir96,%
TamGMS00,%
DelS02}, with the aim of representing large-scale conformation change in terms of a reduced set of low-frequency motions\cite{PetP06}. Another approach is to divide up the protein structure into relatively rigid sections or domains, connected together by flexible regions or ``hinges''. This can be done using a variety of structure-based approaches
%[ JEJ to add references here for 6,36-39 of Gohlke04,
\cite{VihTR94,%6
HolS94,%36
ZehR86,%37
KarS85,%38
MaiA97,%39}
%also {\sc First}/{\sc Froda} hingefinder paper (S Flores, N Echols, D Milburn, BM Hespenheide, K Keating, J Lu, SA Wells, EZ Yu, MF Thorpe and M Gerstein
%"The database of macromolecular motions: new features added at the decade mark." Nucleic Acid Research 34, D296-D301 (2005) ), add to bibliography and cite here. -SAW ].
%\cite{
FloEMH05}.

In this paper we concern ourselves with the ``pebble game''\cite{JacT95}, an integer algorithm for rigidity analysis. By matching degrees of freedom against constraints, it can rapidly divide a network into rigid regions and floppy ``hinges'' with excess degrees of freedom.
%The algorithm is applicable to protein crystal structures if these are treated as molecular frameworks in which bond lengths and angles are constant but dihedral angles may vary; this application, and t
The program {\sc First} implements this algorithm for protein crystal structures\cite{JacRKT01}. The rigid units in a protein structure may be as small as individual methyl groups or large enough to include entire protein domains containing multiple secondary-structure units. The division of a structure into rigid units is referred to as a Rigid Cluster Decomposition (RCD).

Rigidity analysis has been used to study phenomena such as virus capsid assembly \cite{HesJT04} and protein folding \cite{RadHKT02,HesRTK02}. The coarse-graining provided by a RCD also forms the basis of simulation methods aiming to explore the large-amplitude flexible motion of proteins:  the {\sc ROCK} algorithm \cite{ThoLRJ01} and more recently the {\sc Froda} geometric simulation algorithm \cite{WelMHT05}, which has been applied in various studies of protein flexibility \cite{JolWHTF06,HemYB06,JolWFT08,MacNBC07,SunRAJ08}, and the rigidity-enhanced elastic network model \cite{GohT06}.

%The coarse-graining provided by the rigidity analysis dramatically reduces the computational cost of simulating flexible motion while retaining all-atom steric detail, making geometric simulation a promising complement to other methods such as molecular dynamics (MD).

The results of rigidity analysis on proteins depend upon the set of constraints that are included, with the user setting an energy ``cutoff'' which determines the set of hydrogen bonds to include in the analysis,(see section \ref{sec-method}). However, previous studies using {\sc First} have used widely differing, sometimes contradictory, cutoff values and methods of constraint selection --- we give a brief review of the situation in \ref{sec-previous}. This methodological issue not only makes it more difficult for scientists to adopt pebble-game rigidity analysis as a method, but also raises issues in the interpretation of results. There is at present no clear guidance on the ``correct'' choice of cutoff value; nor is it clear how comparable are the results of rigidity analysis using a given cutoff value on slightly different protein structures.

Hence the primary motivation for our study is to fill this gap by explicitly comparing the results of rigidity analysis on groups of very similar crystal structures. We concentrate particularly on eukaryotic cytochrome C while also considering five other proteins (hemoglobin, myoglobin, $\alpha$-lactalbumin, trypsin and HIV-1 protease). For each protein structure we observe the pattern of rigidity loss during the progressive removal of hydrogen bonds, or ``rigidity dilution''\cite{RadHKT02,HesRTK02}. We define {\it mainchain rigidity} as a measure of the rigidity of the protein backbone in order to describe the rigidity loss during dilution. On the basis of this study we comment on the selection of cutoff values and the interpretation of rigidity analyses.

The second motivation for our study is to observe the pattern of rigidity loss during dilution. Previous studies on protein folding \cite{RadHKT02} have drawn comparisons between the folding transition of proteins and the rigidity transition of glassy networks. A recent study \cite{SarWHT07} found that the rigidity transition in glasses could display either first-order or second-order behaviour depending on the character of the constraint network. In the first case, a small change in the constraints causes a sudden transition from an entirely floppy state to one in which the entire system becomes rigid. In the second, rigidity develops in a percolating rigid cluster which initially involves only a small proportion of the network and then gradually increases in size as more constraints are introduced. Our data on rigidity dilution shows that both types of transition are possible in proteins, with four of our proteins typically displaying ``gradual'' rigidity change and two (trypsin and HIV-1 protease) displaying ``sudden'' rigidity change.

%%%%%%%%%%%%%%%%%%%%%%%%%%%%%%%%%%%%%%%%%%%%%%%%%%%%%%%%%%%%%%%%%%%%%%%%
\section{Materials and Methods}
\label{sec-method}
%%%%%%%%%%%%%%%%%%%%%%%%%%%%%%%%%%%%%%%%%%%%%%%%%%%%%%%%%%%%%%%%%%%%%%%%

%%%%%%%%%%%<%%%%%%%%%%%%%%%%%%%%%%%%%%%%%%%%%%%%%%%%%%%%%%%%%%%%%%%%%%%%%
\subsection{Protein selection}
\label{sec-select}

%Previous studies using rigidity analysis on proteins have generally involved either the comparison of no more than two crystal structures of the same protein \cite{HesJT04} or a survey across multiple different proteins, taking a single example of each \cite{RadHKT02}.
%In this study we take a deliberately comparative approach by first choosing a set of proteins, and then, for each protein in the set, obtaining multiple crystal structures.

We have chosen sets of proteins from the protein data bank (PDB)\cite{BerWFG00} to obtain similar crystal structures for our comparison, as summarised in Table \ref{Protein-Summary}. We sought particularly (i) examples of the same protein from different organisms, e.g.\ cytochrome C proteins from multiple different eukaryotic mitochondria, and (ii) protein structures obtained under different conditions of crystallisation, e.g.\ in complex with different ligands, proteins or substrates.
In the present study we will only investigate non-membrane proteins becuase the default treatment of hydrogen bonds and hydrophobic tethers in {\sc First} is based on the assumption that the protein exists in a polar solvent (cytoplasm) rather than being within a hydrophobic or amphiphilic environment as for membrane-bound proteins. Proteins in a membrane environment can still be handled but this requires hand-editing of the constraint network. Rigidity analysis is best carried out on crystal structures with high resolution, so that we can have confidence in the accuracy of the atomic positions when constructing the hydrogen-bond geometries. We therefore concentrated on X-ray crystal structures with resolutions of better than $2.5$\AA.

From each PDB crystal structure we extracted a single protein chain, eliminating all crystal water molecules, but retaining important hetero groups such as the porphyrin/heme units of cytochrome C and hemoglobin. The {\sc PyMOL} visualisation software \cite{Del____} proved very useful for this purpose. We add the hydrogens that are absent from X-ray crystal structures, using the {\sc Reduce} software \cite{WorLRR99} which also performs necessary flipping of side chains. After the addition of hydrogens we renumbered the atoms using {\sc PyMOL} again to produce files usable as input to {\sc First} \cite{JacRKT01,RadHKT02}. In the case of HIV protease we analysed the homodimer unit, as in \cite{JacRKT01}.

%-----------------------------------------------------------------------
\begin{table}
\caption{\label{Protein-Summary} List of all the proteins, their organism of origin, PDB codes as well as the Figures in which they appear}
\begin{center}
\lineup
%\item[]\begin{tabular}{ l l l l }
{\scriptsize\begin{tabular}{ l l l l l }
\br

Protein & Organism & PDB ID & Figure & Comments \cr
 &  &  & \cr
Cytochrome C & Horse & 1HRC & \ref{fig-cyto-en-horse1} & uncomplexed \cr %Monomer\cr
             &       & 1WEJ &  & complexed with antibody E8\cr
             &       & 1U75 &  & complexed with peroxidase\cr%Zinc-Porphyrin Substituted Cytochrome c Peroxidase\cr
             &       & 1CRC &  & at low ionic strength \\ \hline
 & & \cr
%--- & --- & --- \cr
%\marginpar{The cyc are presented in three different graphs}
Cytochrome C & Tuna & 5CYT & \ref{fig-cyto-en-tuna1} & ferricytochrome \cr
             &      & 1I54 &  & 2FE:1ZN mixed-metal porphyrins \cr
             &      & 1I55 &  & 2ZN:1FE mixed-metal porphyrins \cr
             &      & 1LFM &  & Cobalt(III)-subsituted\\  \hline
 & & \cr
%--- & --- & --- \cr
Cytochrome C & Rice & 1CCR & \ref{fig-families}a & \cr%Monomer\cr
             & Bonito & 1CYC & &  \cr%Dimer\cr
             & Bacteria & 1A7V & &  \cr%Dimer (RHODOPSEUDOMONAS PALUSTRIS)  \cr
             & Tuna & 1I55 & &  \cr%Dimer with 2ZN:1FE MIXED-METAL PORPHYRINS\cr
             & Yeast & 1YCC & &  \cr%Monomer\cr
             &       & 2YCC &  & \cr%Monomer\\
\hline

 & & \cr
Myoglobin & Horse & 1DWR & \ref{fig-families}b & \cr%Monomer (horse heart - complexed with CO \cr
          & Whale & 1HJT & &  \cr%Monomer (whale sperm - Ferrous, nitric oxide bound\cr
          & Turtle & 1LHS &  & \cr%Monomer (loggerhead sea turtle)\\
\hline
 & & \cr
$\alpha$-lactalbumin & Baboon & 1ALC & \ref{fig-families}c & \cr%Monomer \cr
                     & Human & 1HML &  & \cr%Monomer \cr
                     & Goat & 1HFY &  & \cr%Dimer \cr
                     & Human & 1A4V &  & \cr%Monomer \cr
                     & Guinea pig & 1HFX &  & \cr%Monomer \cr
                     & Cattle & 1F6R &  & \cr%Hexamer\\
\hline
 & & \cr
Hemoglobin & Human & 1A3N & \ref{fig-families}d & deoxy \cr%Dimer\cr
($\alpha$ chain)       &  & 2DN1 & &  oxy\cr%Dimer Oxy \cr
           &       & 2DN2 & &  deoxy\cr%Tetramer Deoxy\cr
           &       & 2DN3 &  & carbonmonoxy\cr%Dimer Carbonmonoxy\\
           & Goose & 1A4F & &  \cr%Oxy \cr%Monomer \cr
           & Rice & 1D8U &  & \cr%Dimer \cr
           & Bacteria & 1DLW &  & \cr%Monomer - TRUNCATED HEMOGLOBIN ! \cr
           & Alga & 1DLY &  & \cr%Monomer (Green unicellular alga)\cr
           & Cattle & 1G09 &  & \cr%Carbonmonoxy\cr%Tetramer Carbonmonoxy  \cr
           & Worm & 1KR7 &  & \cr%Monomer Nemertean worm Cerebratulus lacteu \cr
           & Clam & 1MOH &  & \cr%Monomer \cr

\hline

 & & \cr
%HIV-1 Protease & Virus & 3PHV & Monomer (wild type) \cr
% (Figure \ref{fig-families}e) &       & 1HHP & Monomer (wild type) \cr
HIV-1 Protease & Virus & 1HTG & \ref{fig-families}e & homodimers with inhibitors bound \cr%(wild type) \cr
%               &       & 3HVP & Monomer (Synthetic type) \cr
  &       & 4HVP & & \cr%Dimer with inhibitor (Synthetic type) \cr
               &       & 7HVP & &\cr%Dimer with inhibitor (Synthetic type) \cr
               &       & 8HVP & &  \cr%Dimer with inhibitor (Synthetic type) \cr
               &       & 9HVP & &  \cr%Dimer with inhibitor (wild type) \cr
\hline

 & & \cr
Trypsin & Salmon & 1A0J & \ref{fig-families}f & \cr%Tetramer\cr
        & Cattle & 1AQ7 & \cr%Monomer with inhibitor Aeruginosin 98-B\cr
        &        & 1AUJ & \cr%Monomer\cr
        & Pig    & 1AVW & \cr%Monomer with soya bean trypsin inhibitor (Orthorhombic crystal form)  \cr
        & Pig    & 1AVX & \cr%Monomer with soya bean trypsin inhibitor (Tetragonal crystal form) \cr
        & Cattle & 1AZ8 & \cr%Monomer with bis-phenylamidine inhibitor\cr
        & Rat    & 1BRA & \cr%Monomer \cr
        &        & 1BRB & \cr%Monomer anionic trypsin with inhibitor BPTI \cr
        &        & 1BRC & \cr%Monomer anionic trypsin with inhibitor APPI\cr
        & Cattle & 1BTH & \cr%THROMBIN Dimer with BPTI **** DELETE THIS ENTRY?\cr
        & Salmon & 1BZX & \cr%Monomer with BPTI  \cr
        & Human  & 1H4W & \cr%Monomer (Brain Trypsin) \cr
        &        & 1HPT & \cr%PANCREATIC SECRETORY TRYPSIN INHIBITOR (KAZAL TYPE) VARIANT 3!!\cr
        & Cattle & 1K1I & \cr%Monomer with inhibitor\cr
        &        & 1K1J & \cr%Monomer with inhibitor\cr
        &        & 1K1M & \cr%Monomer with inhibitor\cr
        &        & 1K1N & \cr%Monomer with inhibitor\cr
        &        & 1K10 & \cr%Monomer with inhibitor\cr
        &        & 1K1P & \cr%Monomer with inhibitor\cr
        & Pig    & 1LDT & \cr%Monomer with leech-derived Tryptase inhibitor\cr
        & Human  & 1TRN & \cr%Dimer \cr
        &        & 2RA3 & \cr%Dimer (cationic trypsin) with BPTIs\cr
        & Rat    & 3TGI & \cr%Monomer (Anionic Trypsin) with BPTI\\ \hline

\end{tabular}
}
\end{center}
\end{table}
%-----------------------------------------------------------------------

%%%%%%%%%%%%%%%%%%%%%%%%%%%%%%%%%%%%%%%%%%%%%%%%%%%%%%%%%%%%%%%%%%%%%%%%
\subsection{Rigidity analysis and dilution}
\label{sec-software}

% %-----------------------------------------------------------------------
% \begin{figure}[tb]
% \begin{center}
% \includegraphics[width=0.6\textwidth, angle=0]{hbonddistance.eps}
% \end{center}
% \caption{Dependence of hydrogen bond energy $E$ as used in {\sc First} on the donor-acceptor distance. The shaded region indicates how an distance variation of $\pm 0.1$\AA\ can lead to a variation in the bond energy of more than $1$ kcal/mol. \label{fig-hbdist}}
% \end{figure}
% %-----------------------------------------------------------------------

\begin{figure}[tb]

\begin{center}
\includegraphics[width=0.6\textwidth, angle=0]{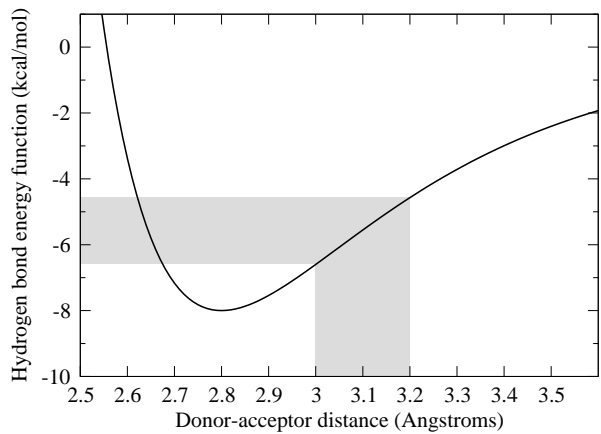}
\end{center}
\caption{Dependence of hydrogen bond energy $E$ in {\sc First} on the donor-acceptor distance. The shaded region indicates how an distance variation of $\pm 0.1$\AA\ can lead to a variation in the bond energy of more than $1$ kcal/mol. \label{fig-hbdist}}
\end{figure}

%, which was run over the resulting processed PDB file with appropriate command line options. \footnote{{options ``-E 0 -dil 1 -non''}. }
The energy of each potential hydrogen bond in the processed structure is calculated in {\sc First} using the Mayo potential \cite{DahGM97}; the distance-dependent part of this potential is shown in Figure \ref{fig-hbdist}. For the dilution, {\sc First} performs an initial rigidity analysis including all bonds with energies of $0$ kcal/mol or lower; bonds are then removed in order of strength, gradually reducing, or ``diluting'', the rigidity of the structure.

%-----------------------------------------------------------------------
%-----------------------------------------------------------------------

%%              1HRC 3D
%-----------------------------------------------------------------------
\begin{figure}[ptb]
%\label{fig-horse-stripy}
\begin{center}
\includegraphics[width=0.8\textwidth]{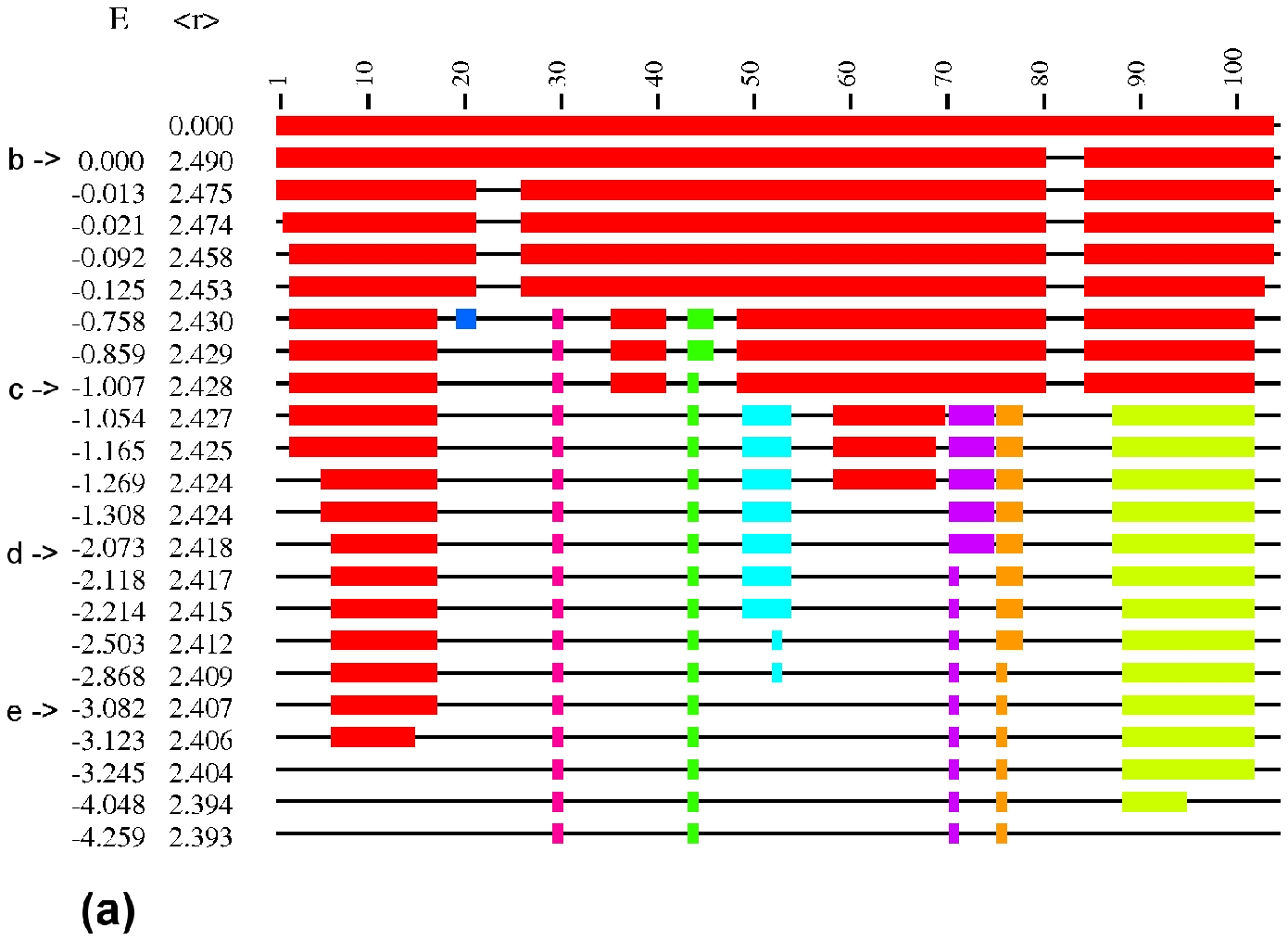}

\includegraphics[width=0.8\textwidth]{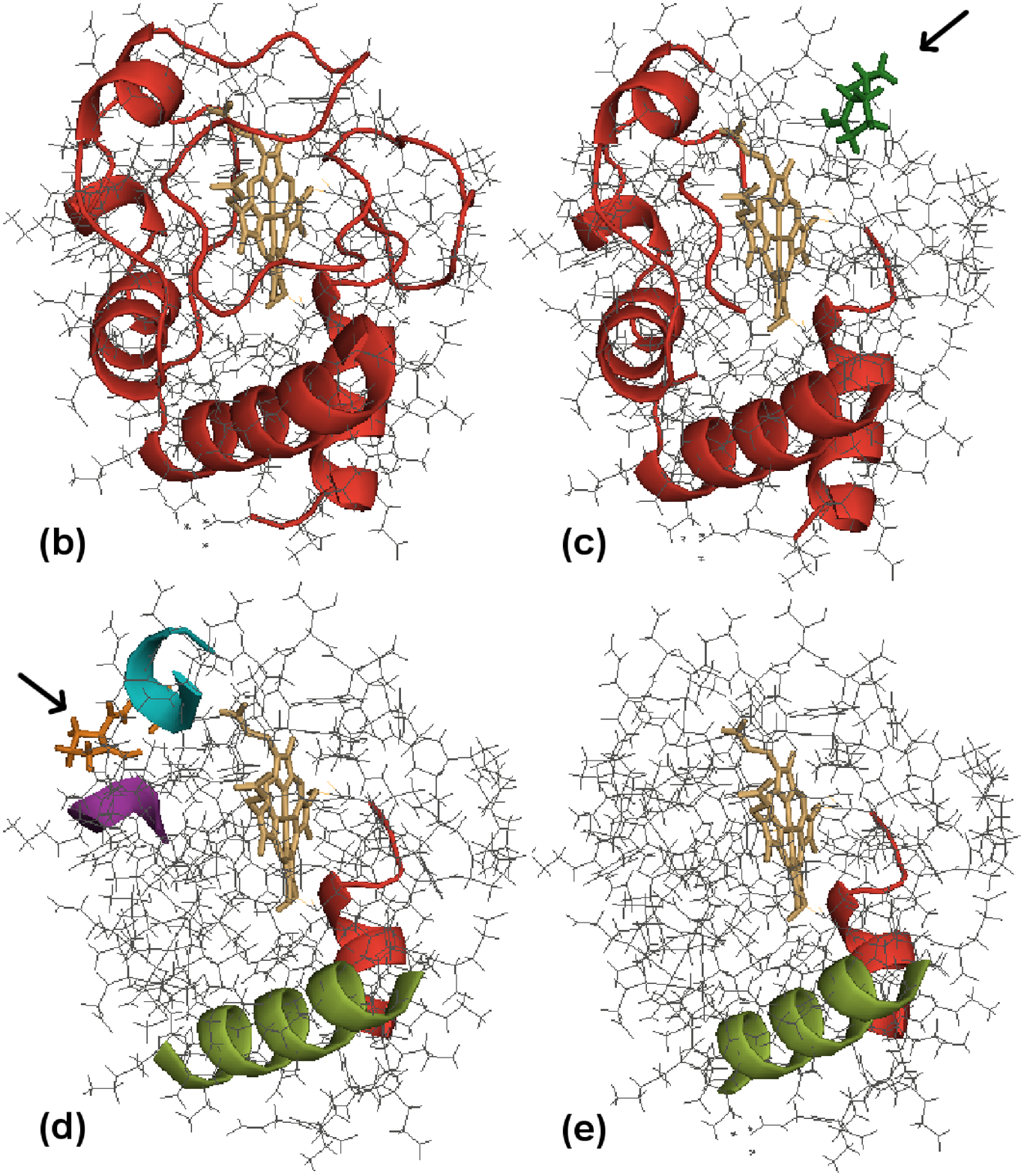}

%1HRC_FOUR_TEST.eps}

% \includegraphics[width=0.45\textwidth]{1HRC_0000.eps}
% \includegraphics[width=0.49\textwidth]{1HRC_1007.eps}
% $2.073$\includegraphics[width=0.4\textwidth]{1HRC_2073.eps}
% $0.000$ $1.007$ $3.082$\includegraphics[width=0.4\textwidth]{1HRC_3082.eps}
\end{center}

\caption{(a) Dilution plot for horse cytochrome C from the 1HRC structure. Flexible regions of the polypeptide chain appear as black thin lines, whereas rigid portions appear as coloured along the protein chain with C$_\alpha$ labelled from 1 to 105. The second colum on the left indicates the mean number $\langle r\rangle$ of bonded neighbours per atom as the energy cutoff $E$ changes. When $E$ decreases (left-most column), rigid clusters break up and more of the chain becomes flexible. Colour coding shows which atoms belong to which rigid cluster. 
(b,c,d and e) Rigidity distribution for horse Cytochrome C from the 1HRC structure in 3D. These figures represent in grey the flexible regions and in colour the largest 
%1, 2, 5 and 2 
rigid regions for the native state at energy cutoffs (b) $E=0.000$, (c) $E=1.007$, (d) $E=2.073$ and (e) $E=3.082$, respectively. For each figure, the colour coding correlates with the colour coding given in (a). The arrows in (c) and (d) indicate two smaller rigid clusters shown in ``stick'' representation for clarity.
The heme group is shown in ``stick'' representation (yellow). 
 \label{3D}}
\end{figure}
%-----------------------------------------------------------------------

An example of this rigidity dilution for a given protein is shown in Figure \ref{3D}a for the 1HRC horse cytochrome C structure. The horizontal axis represents the protein's linear primary structure. Flexible areas  of the polypeptide sequence are shown as horizontal thin black lines while areas lying within a rigid cluster are shown as thicker coloured blocks. Colour is used to differentiate which residues belong to which rigid cluster. The three-dimensional protein fold makes it possible for residues that are widely separated along the backbone to be spatially adjacent and form a single rigid cluster. The vertical axis on the dilution plot represents the dilution of constraints by progressively lowering the cutoff energy for inclusion of hydrogen bonds in the constraint network. Each time the rigid cluster analysis of the mainchain $\alpha$-carbon atoms (C$_\alpha$) changes as a result of the dilution, a new line is drawn on the plot, labelled with the energy cutoff and with the network mean coordination for the protein at that stage. We should stress that the RCD is always performed over the entire protein structure (mainchain and sidechain atoms) and a dilution is performed for every hydrogen bond removed from the set of constraints, typically several hundred bonds for a small globular protein. The dilution plot is then a summary concentrating on the rigid-cluster membership of the C$_\alpha$ atoms defining the protein backbone.

%%%%%%%%%%%%%%%%%%%%%%%%%%%%%%%%%%%%%%%%%%%%%%%%%%%%%%%%%%%%%%%%%%%%%%%%
\subsection{Mainchain rigidity loss during dilution}
\label{sec-tracking}

%-----------------------------------------------------------------------
\begin{figure}[tb]
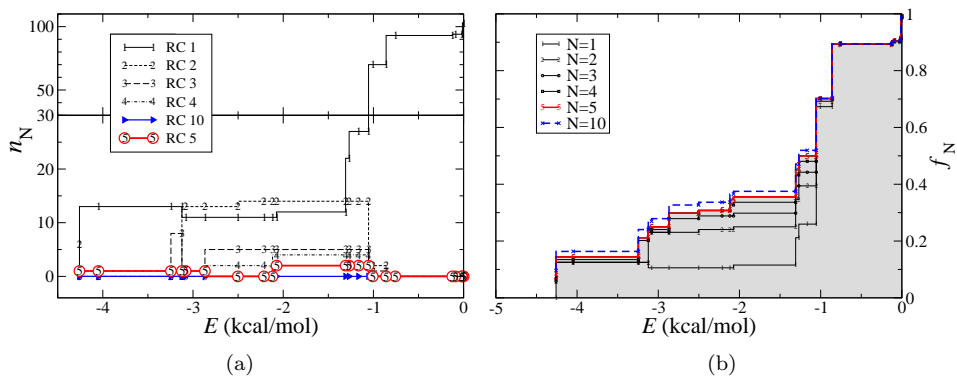

\begin{center}

\subfigure[]%[Membership in larger rigid clusters]
{\includegraphics[scale=0.25]{Figure3a.eps} }
\subfigure[]%[Fraction of mainchain belonging to large rigid clusters]
{\includegraphics[scale=0.25]{Figure3b.eps} }
\caption{(a) The number $n_N$ of C$_\alpha$ atoms contained within rigid clusters (RC) $N=1,\ldots,5$ and $10$ of the 1HRC structure.
% Lines are for $N=1, \ldots, 5$ and $10$ and are labelled by the number of the rigid clusters.
 Smaller, higher-numbered clusters do not contain more than one C$_\alpha$.
 (b) The fraction $f$ of the protein's C$_\alpha$ atoms contained within clusters $1$ to $N$. The line corresponding to the $N=5$ data has been shaded to show that the inclusion of rigid clusters $1$ through $5$ captures the large-scale rigidity of the protein.
}
\label{fig-track-count}
\label{fig-track-frac}
\end{center}
\end{figure}
%-----------------------------------------------------------------------

Dilution plots of very similar protein structures can be compared directly as shown in Figure \ref{fig-horse-stripy}. This form of comparison, however, becomes unwieldy when comparing large numbers of structures, and can obscure differences in the hydrogen-bond energy scale. For glassy networks \cite{SarWHT07} the overall degree of rigidity of the structure was measured by the number of atoms in the largest spanning rigid cluster in a network with periodic boundary conditions.
%We therefore extract from the dilution plots a quantity measuring the overall degree of rigidity of the structure, and plot it as a function of the hydrogen-bond energy cutoff $E$.
Since the protein is not a periodic structure, we measure its overall rigidity by considering how many of its residues are included in large rigid clusters.
%
%For protein structures, we first extract the number $n_N(E)$ of C$_\alpha$ belonging to each of the first $N$ largest rigid clusters. Then we normalize $n_N(E)$ as a fraction the total number ${\cal N}_{\rm CA}$ of C$_\alpha$. This allows us to compare the rigidity between different proteins.
%
%So our measure of overall rigidity $f_N(E)$ is the fraction of atoms that are found in the $N$ largest rigid clusters,
%\begin{equation}
% f_N(E) = n_N(E)/{\cal N}_{\rm CA} \quad
%\end{equation}
%(essentially, those that appear as large blocks in the dilution plot).

In Figure \ref{fig-track-count}(a) we show the number $n_N$ of C$_\alpha$ contained within the larger $N$ rigid clusters of the horse cytochrome C structure 1HRC, for which the total number of C$_\alpha$ atoms equals ${\cal N}_{{\rm C}_\alpha}=105$. It is clear that only the first few rigid clusters (numbered $1$--$5$) contain more than one C$_\alpha$ while higher-numbered clusters do not contain more than one C$_{\alpha}$ and do not represent two or more residues forming a single rigid unit.
%Typically single residues containing one C$_\alpha$ exist as rigid clusters when enough hydrogen bonds have been removed from the system. This happens for example with proline due to its cyclised structure.
In Figure \ref{fig-track-frac}(b) we show the fraction $f_N$ of C$_\alpha$ contained in the first $N$ cluster, defined as
\begin{equation}
 f_N(E) = \frac{1}{\cal N}_{{\rm C}_\alpha} \sum_1^N n_N(E)
\end{equation}
for, e.g.\ those C$_\alpha$ lying within rigid clusters $N=1$ to $5$ and also $10$. 
The inclusion of the first five rigid clusters captures the large-scale rigidity of the protein; the difference between $N=5$ and $N=10$ is minimal. We therefore use the $N=5$ measure, $f_5(E)$, to quantify protein rigidity hereafter, which we will refer to as {\it mainchain rigidity}. We emphasize that we have also computed all results presented here for $N=4$ and $N=6$ with quantitatively similar and qualitatively identical results.

It is worth noting the "stepped" appearance of our graphs. This is because a given pattern of rigidity persists as the cutoff is lowered until at a specific value it changes and a certain amount of rigidity is lost.

\subsection{Structural comparison by RMSD}

When dealing with slightly varying crystal structures of the same protein, we quantify the structural variation by aligning the C$_\alpha$ atoms of two structures 
%\footnote{Using the {\sc PyMOL} ``align'' command \cite{Del____} }
and obtaining the root-mean-square deviation between C$_\alpha$ positions,
\begin{equation}
d = \sqrt{ \frac{1}{N_{{\rm C}_\alpha}} \sum_{i=1}^{N_{{\rm C}_\alpha}} d_{ii}^2 } 
\end{equation}
where $d_{ii}$ is the distance between the C$_\alpha$ atoms of residue $i$ in the aligned structures.

%%%%%%%%%%%%%%%%%%%%%%%%%%%%%%%%%%%%%%%%%%%%%%%%%%%%%%%%%%%%%%%%%%%%%%%%
\section{Results and discussion}
\label{sec-results}
%%%%%%%%%%%%%%%%%%%%%%%%%%%%%%%%%%%%%%%%%%%%%%%%%%%%%%%%%%%%%%%%%%%%%%%%

%%%%%%%%%%%%%%%%%%%%%%%%%%%%%%%%%%%%%%%%%%%%%%%%%%%%%%%%%%%%%%%%%%%%%%%%
\subsection{Comparing rigidity of very similar proteins: cytochrome C}
\label{sec-results-cytC}
%-----------------------------------------------------------------------
\begin{figure}[tb]
%\label{fig-horse-stripy}
\begin{center}
  \subfigure[1HRC: uncomplexed]{\includegraphics[scale=0.47]{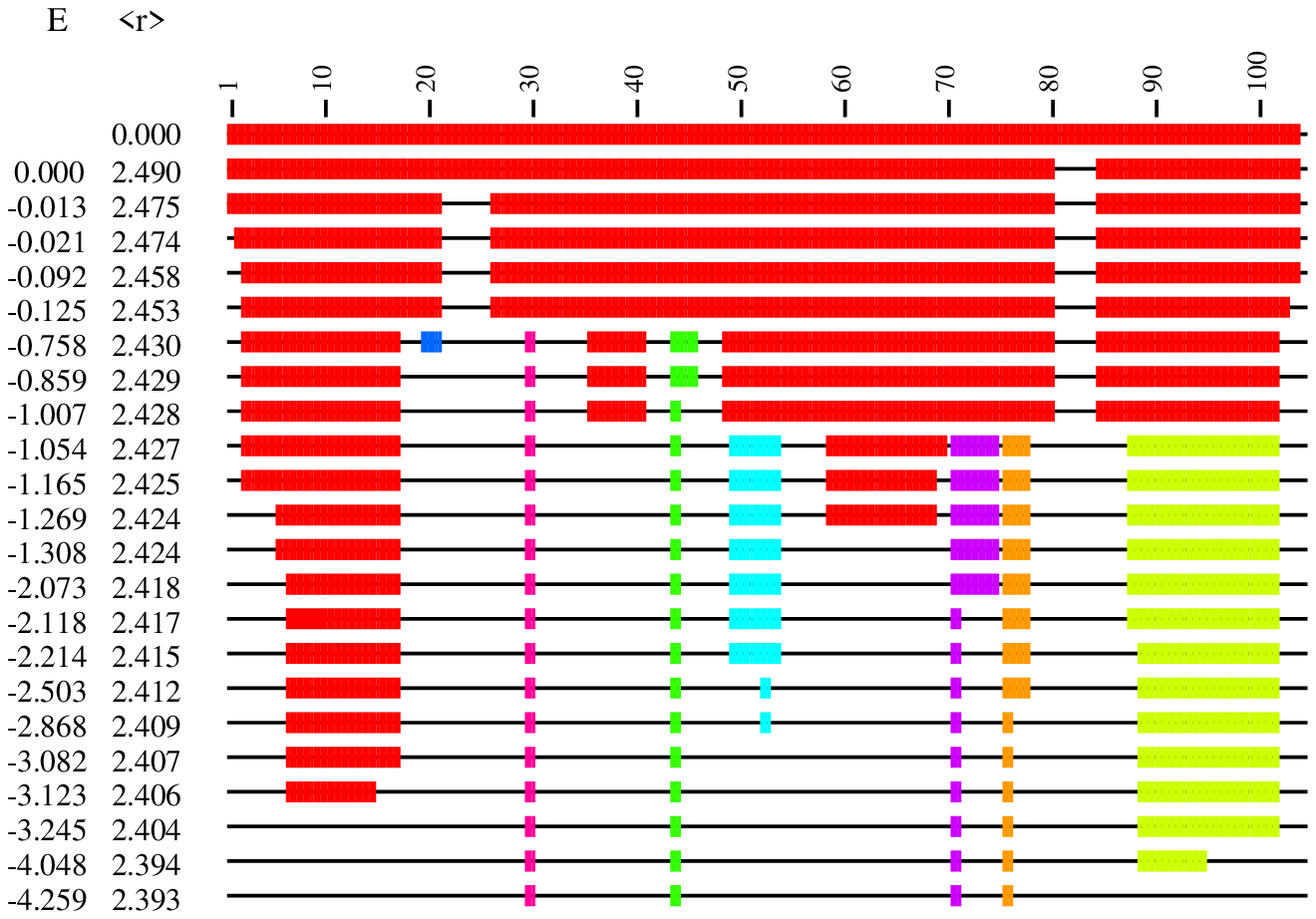} }
  \subfigure[1WEJ: antibody complex]{\includegraphics[scale=0.47]{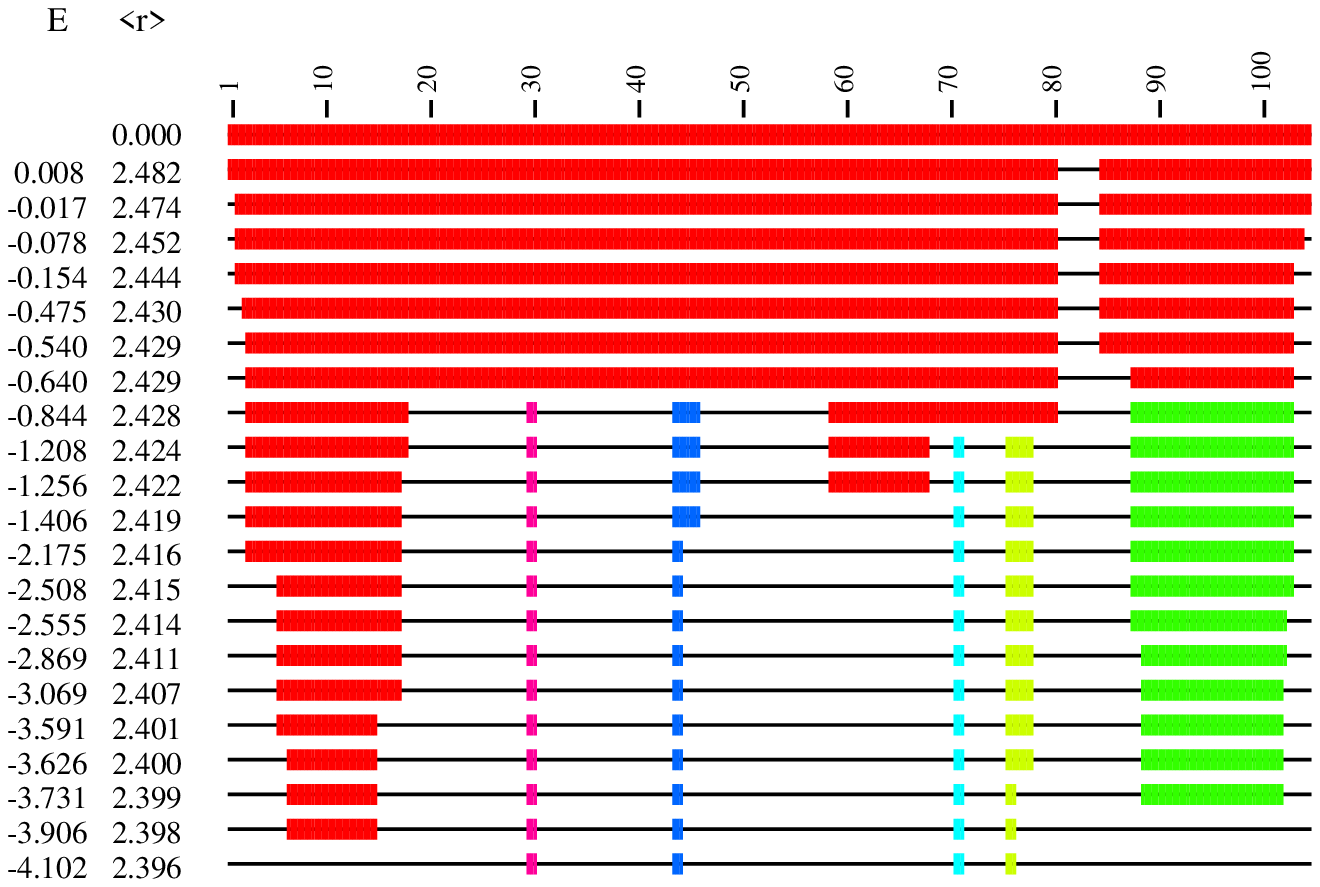} }
  \subfigure[1U75: peroxidase complex]{\includegraphics[scale=0.47]{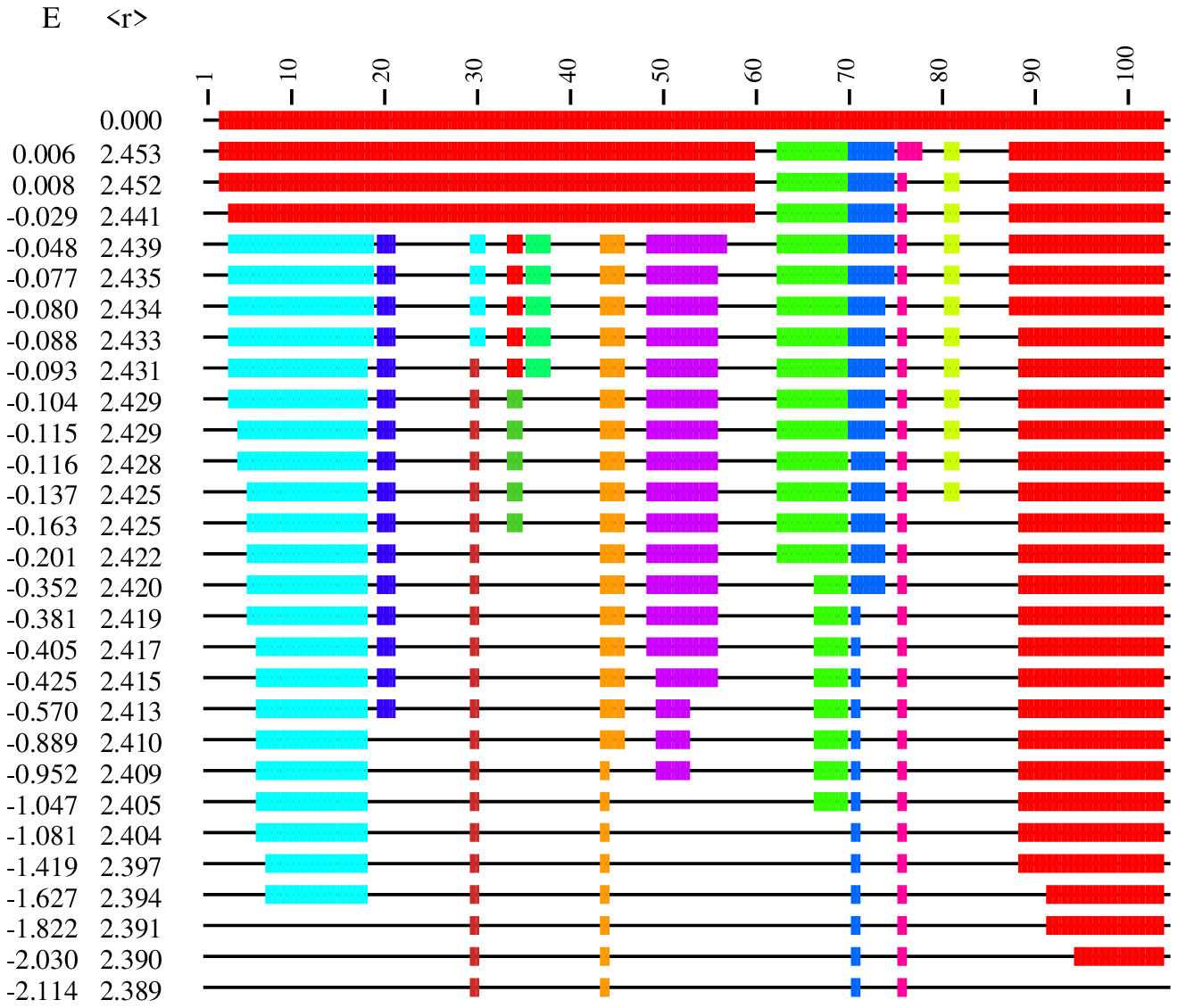} }
  \subfigure[1CRC: low ionic strength]{\includegraphics[scale=0.47]{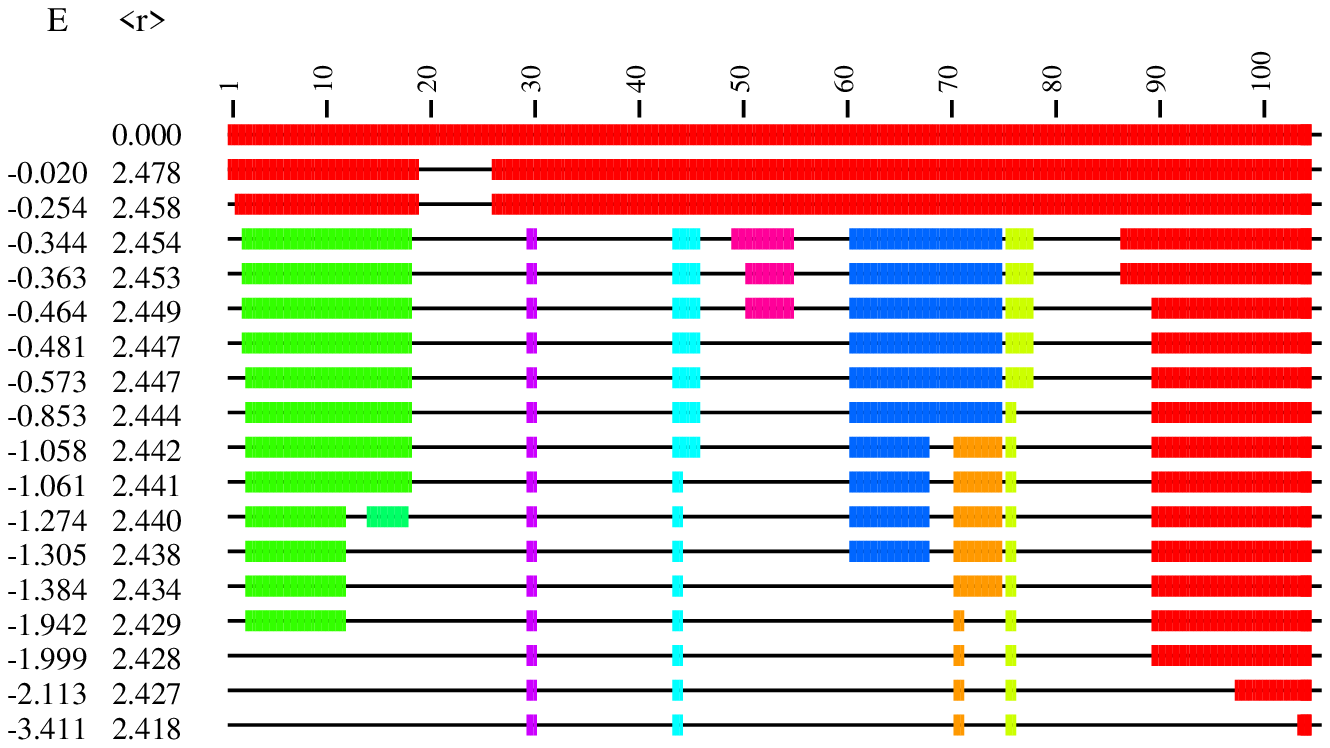} }
\caption{Dilution plots for four crystal structure of horse cytochrome C. The four structures are very similar to each other (see text) and display similar patterns of rigidity loss. The central portion of the protein sequence breaks up into smaller clusters (e.g. close to E=-1 for 1HRC and E=-0.7 for 1WEJ) and then becomes entirely flexible, while the rigidity of the two ends of the sequence, around residues 5--15 and 90--105, persists longer; these portions are $\alpha$-helical in secondary structure.}
 \label{fig-horse-stripy}
\end{center}
\end{figure}
%-----------------------------------------------------------------------

%-----------------------------------------------------------------------
\begin{table}[tb]
%\caption{\label{tab-horse-1} RMSD variations in $\alpha$-carbon positions among four horse cytochrome C structures, in \AA\ , showing the similarity of the structures.}
(a)\begin{minipage}[tb]{0.48\textwidth}
%\begin{indented}
%\lineup
%\item[]
\begin{tabular}{ c c c c }
\br
From$\backslash$To:&1HRC&1CRC&1WEJ\cr
1CRC&0.32&---&---\cr
1WEJ&0.318&0.321&---\cr
1U75&0.472&0.53&0.572\cr
\br
\end{tabular}
\end{minipage}
%\end{indented}
%\end{table}
%-----------------------------------------------------------------------
%-----------------------------------------------------------------------
(b)\begin{minipage}[tb]{0.48\textwidth}
%\begin{table}[tb]
%\caption{\label{tab-tuna-1} RMSD variations in $\alpha$-carbon positions among four tuna cytochrome C structures, in \AA.}
%\begin{indented}
%\lineup
%\item[]
\begin{tabular}{ c c c c }
\br
From$\backslash$To:&5CYT&1I55&1I54\cr
1I55&0.27&---&---\cr
1I54&0.2668&0.041&---\cr
1LFM&0.286&0.116&0.087\cr
\br
\end{tabular}
%\end{indented}
\end{minipage}
\caption{Root-mean-square deviation in \AA\ for C$_\alpha$ positions among (a) four horse cytochrome C structures and (b) four tuna cytochrome C structures, showing the similarity of the structures.}
\label{tab-horse-1}
\label{tab-tuna-1}
\label{tab-cytochromeC}
\end{table}
%-----------------------------------------------------------------------

In Figure \ref{fig-horse-stripy} we show dilution plots for four mitochondrial cytochrome C structures obtained from horse crystallised under different conditions as detailed in Table \ref{Protein-Summary}. The structural variations between these four structures are small (Table \ref{tab-horse-1}a), the largest being $0.572$\AA\ between 1U75 and 1WEJ; for comparison, Minary and Levitt \cite{MinL08} consider structures within $d \simeq 4$\AA\ as ``near-native''.

The patterns of rigidity loss in Figure \ref{fig-horse-stripy} appear quite similar on first inspection. The central portion of the protein sequence breaks up into smaller clusters and then becomes entirely flexible, while the rigidity of the two ends of the sequence, around residues 5--15 and 90--100, persists longer; due to this persistence, these portions ($\alpha$-helical in secondary structure) were identified in \cite{HesRTK02} as being the folding core of cytochrome C, in agreement with experimental evidence.

On closer inspection, however, we can see differences between the four structures in the cutoff energies in which changes in rigidity take place. For example, in structures 1HRC and 1WEJ, the terminal $\alpha$-helical sequences remain rigid down to cutoff values below $-3$ kcal/mol, while in 1CRC and 1U75 these sequences are already largely flexible at a cutoff value of $-2$ kcal/mol.
We plot the mainchain rigidity of these four proteins as a function of cutoff energy during dilution in Figure \ref{fig-cyto-en-horse1}(a). The differences in energy scale of the rigidity loss is now clearly visible. Note in particular that in the energy range around $-0.1$ to $-0.6$   kcal/mol, two of the structures retain mainchain rigidity ($ f_5 > 0.9 $) while the other two have already dropped to $f_5 < 0.5$.

%-----------------------------------------------------------------------
\begin{figure}[tb]
\begin{center}
  \subfigure[5CYT: normal ferricytochrome]{\includegraphics[scale=0.44]{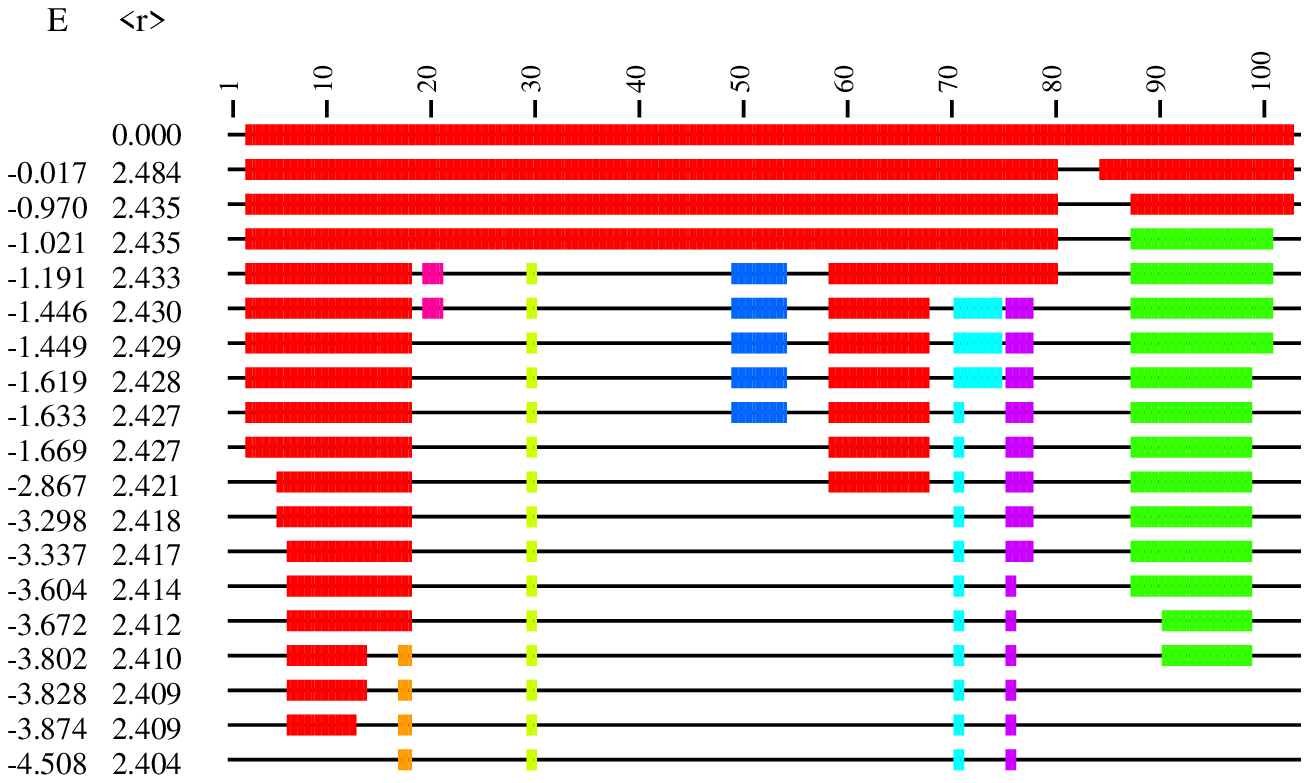} }
  \subfigure[1I55: crystallised from 2Zn:1Fe mix]{\includegraphics[scale=0.44]{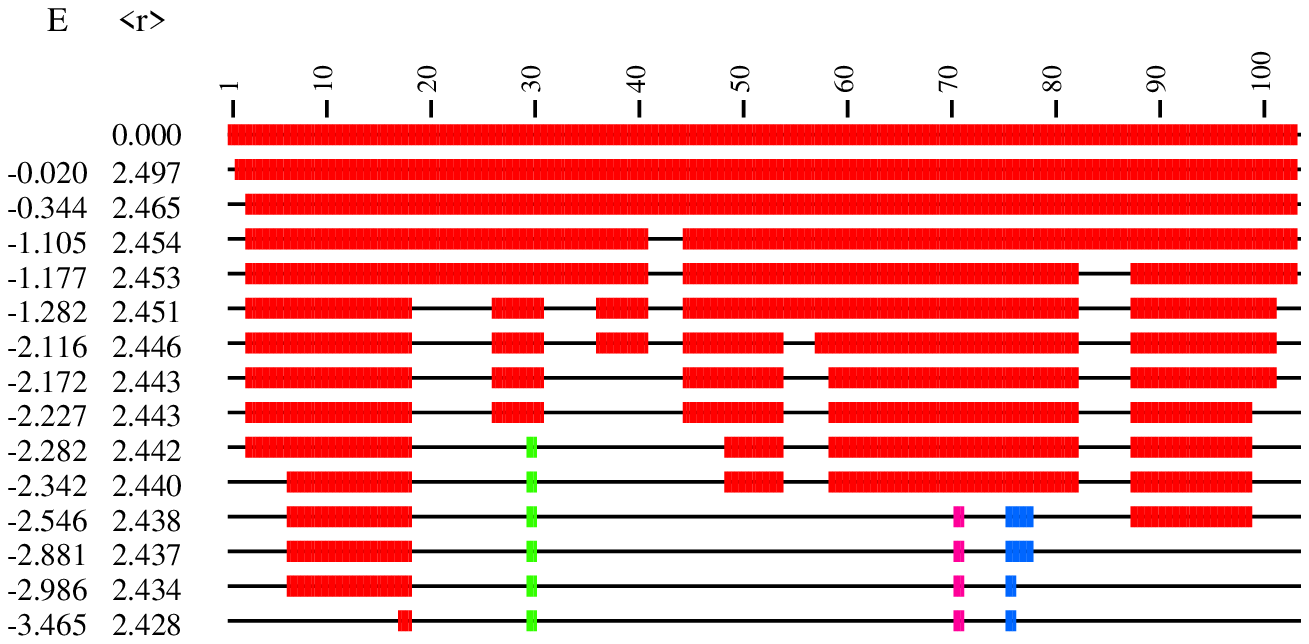} }
  \subfigure[1I54: crystallised from 1Zn:2Fe mix]{\includegraphics[scale=0.44]{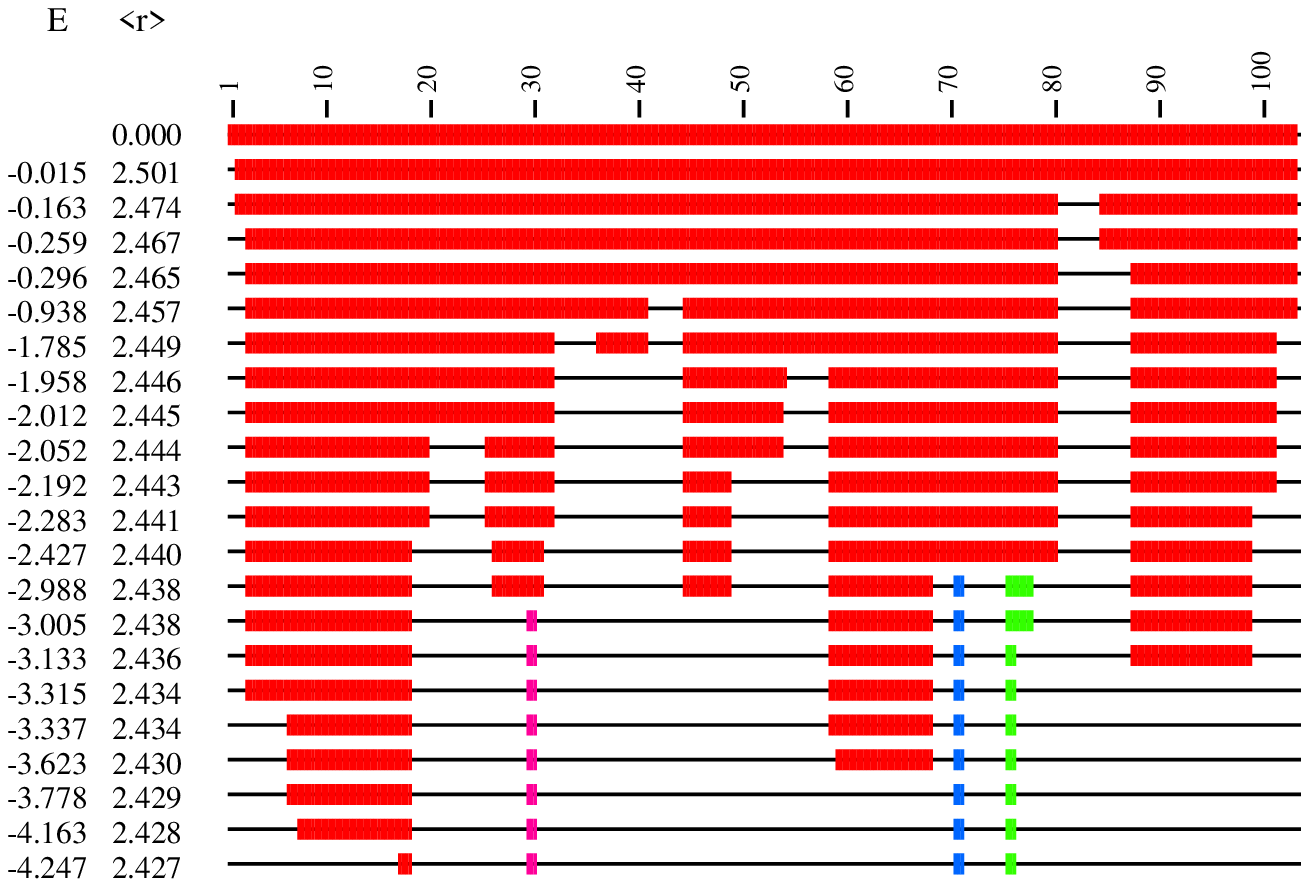} }
  \subfigure[1LFM: with Co replacing Fe]{\includegraphics[scale=0.44]{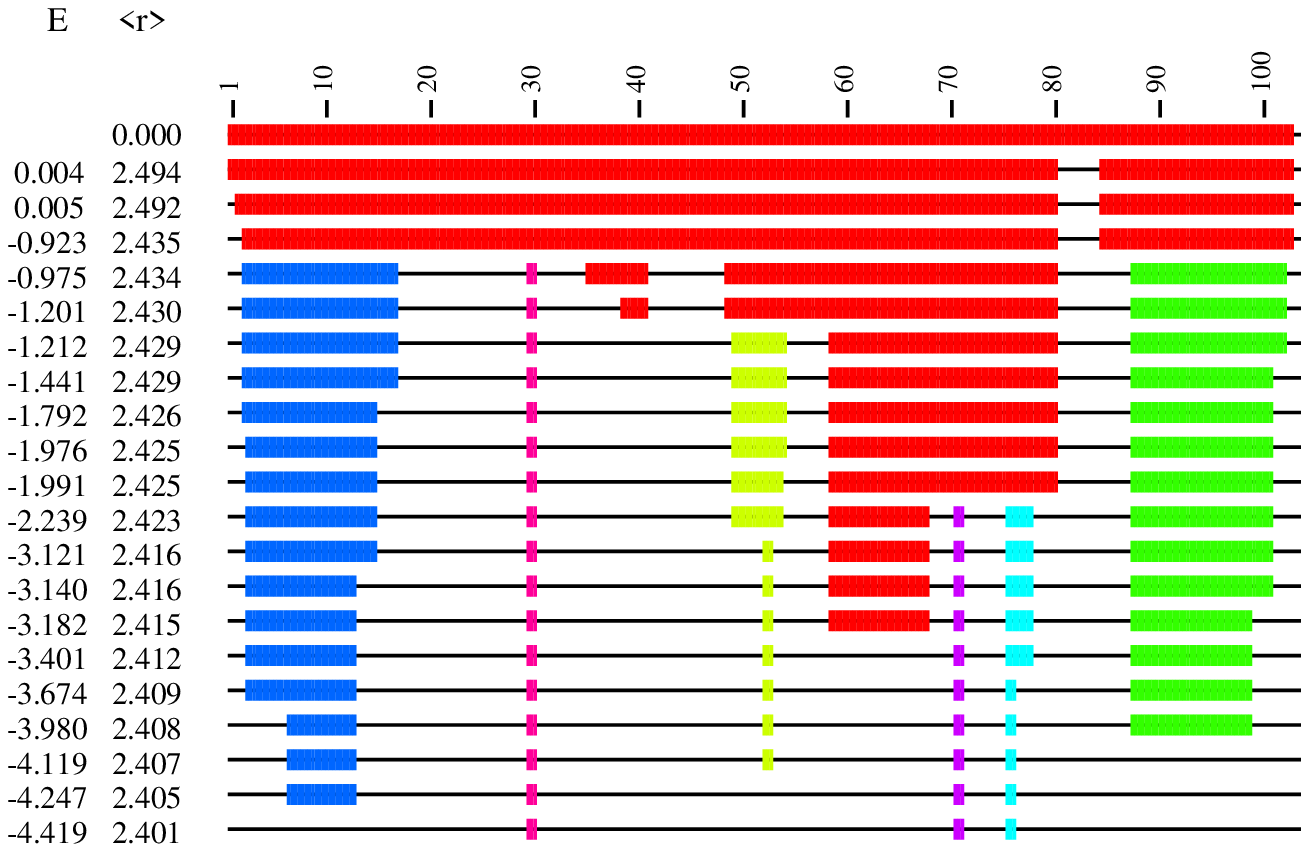} }
\caption{Rigidity dilutions for four forms of tuna cytochrome C crystallised with different metal ion content in the heme groups. (a) normal Fe, (b) from a mixture with 2Zn:1Fe, (c) from a mixture with 2Fe:1Zn, (d) with Co. \label{fig-tuna-stripy}}
\end{center}
\end{figure}
%-----------------------------------------------------------------------

%-----------------------------------------------------------------------
\begin{figure}[t]
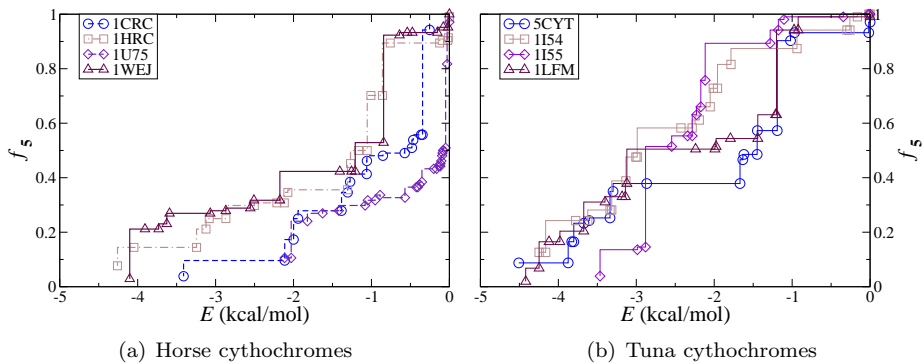

\begin{center}
  \subfigure[Horse cythochromes]{\includegraphics[scale=0.24]{Figure6a.eps} }
  \subfigure[Tuna cythochromes]{\includegraphics[scale=0.24]{Figure6b.eps} }

\caption{ (a) Mainchain rigidity as a function of hydrogen bond energy cutoff $E$ during dilution for four horse mitochondrial cytochrome C structures. Note that for cutoff energy values in the region of $-0.5$ kcal/mol, structure 1HRC and 1WEJ are almost completely rigid while structures 1U75 and 1CRC are less than $50$\%\ rigid. \label{fig-cyto-en-horse1}
(b) Mainchain rigidity for four tuna cytochrome C structures. Note the considerable differences in behaviour between, for example, 5CYT and 1I55 in the $-1$ to $-2$ kcal/mol energy range, even though the structures differ from each other only slightly. \label{fig-cyto-en-tuna1}}
\label{fig-cyto-en}
\end{center}

\end{figure}
%-----------------------------------------------------------------------

In Figure \ref{fig-tuna-stripy}, we now consider mitochondrial cytochrome C structures (from tuna) which differ only in their heme-group metal content and are structurally very similar (RMSD values given in Table \ref{tab-tuna-1}b). We  see that the dilution plots for the tuna protein have similar shapes and indeed are quite similar to those for the horse protein (Figure \ref{fig-horse-stripy}). There are differences, however: in particular, in structure 1I54 the $\alpha$-helical region at residues $60$--$70$ remains rigid to lower cutoff values then that at residues $90$--$100$, which would disagree with the ``folding core'' prediction of reference  \cite{HesRTK02}. We would therefore argue that physical conclusions drawn from rigidity analysis should be based on the comparison of as many structures as possible if they are to be robust.

Once we plot mainchain rigidity as a function of cutoff energy we again observe differences in the energy scales at which rigidity is lost, (Figure \ref{fig-cyto-en-tuna1}b). The greatest discrepancy appears in the energy range from $-1$ to $-2$ kcal/mol; here the 5CYT structure has $f_5 \simeq 0.4$ while 1I55 has $f_5 \simeq 0.9$, although the structures differ by less than $d=0.3$\AA\ in C$_\alpha$ RMSD.

%%%%%%%%%%%%%%%%%%%%%%%%%%%%%%%%%%%%%%%%%%%%%%%%%%%%%%%%%%%%%%%%%%%%%%%%
\subsection{Variability of energy scales and selection of cutoff values}
\label{sec-vary}

It is clear from our investigation of cytochrome C structures that the rigidity analysis at a given cutoff value on very similar structures can easily produce different results. This is not because the dilution plots for these structures differ drastically in their shape, but rather because the cutoff energy at which a major change in rigidity takes place can differ by approximately $1$ kcal/mol between very similar structures. This sensitivity of cutoff energy scales to small structural variations is understandable if we consider, for example, the distance dependence for the hydrogen-bond energy function \cite{JacRKT01}; we show in Figure \ref{fig-hbdist} that a variation in donor-acceptor distance of only $0.1$\AA\ can shift the hydrogen bond energy by around $1$ kcal/mol. Thus while the hydrogen bond energy function is successful in distinguishing weaker from stronger bonds, its resolution is limited to approximately $1$ kcal/mol.

This implies that exact values of the hydrogen bond cutoff energy cannot easily be transferred between different crystal structures. Rather, it is advisable to perform rigidity dilution on the specific protein structure(s) of interest and to observe how the rigidity changes as the weaker bonds are eliminated, and which portions of the structure are most stable, before selecting an appropriate cutoff value for further investigation of the rigidity/flexibility of the structure(s). While this is an a sense the implicit message of the wide variety of cutoff values used in previous studies (see section \ref{sec-previous}) we believe the point should be made explicitly for the benefit of potential users of the method.

%%%%%%%%%%%%%%%%%%%%%%%%%%%%%%%%%%%%%%%%%%%%%%%%%%%%%%%%%%%%%%%%%%%%%%%%
\subsection{Patterns of rigidity loss}
\label{sec-pattern}
%-----------------------------------------------------------------------
\begin{figure}[tb]
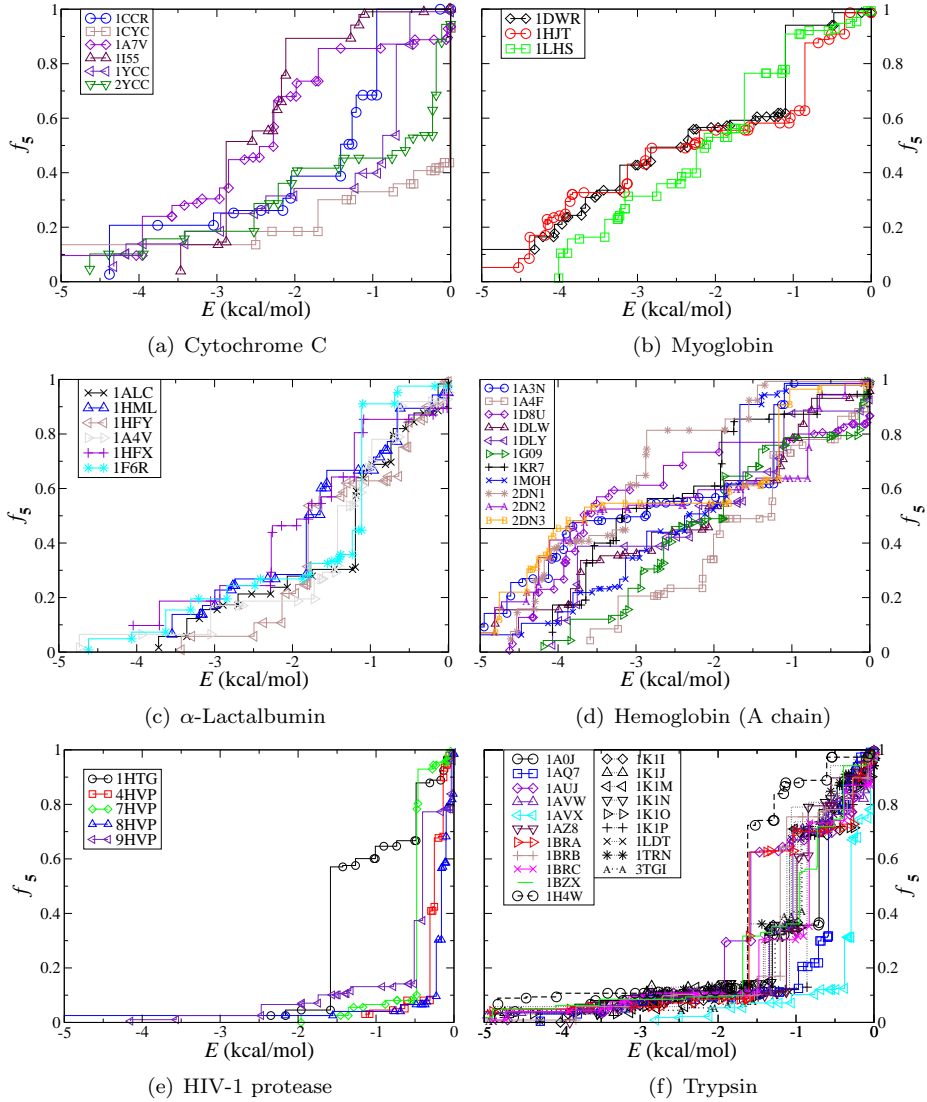


\begin{center}
  \subfigure[Cytochrome C]{\includegraphics[scale=0.24]{Figure7a.eps} }
  \subfigure[Myoglobin]{\includegraphics[scale=0.24]{Figure7b.eps} }
  \subfigure[$\alpha$-Lactalbumin]{\includegraphics[scale=0.24]{Figure7c.eps} }
  \subfigure[Hemoglobin (A chain)]{\includegraphics[scale=0.24]{Figure7d.eps} }
  \subfigure[HIV-1 protease]{\includegraphics[scale=0.24]{Figure7e.eps} }
  \subfigure[Trypsin]{\includegraphics[scale=0.24]{Figure7f.eps} }
\caption{Rigidity dilutions for different families of proteins: cytochrome C, myoglobin, $\alpha$-lactalbumin, hemoglobin, HIV-1 protease and trypsin. We can see that proteins can display either a ``gradual'' (a--d) or a ``sudden'' (e--f) pattern of rigidity loss. \label{fig-families}}
\end{center}
\end{figure}
%-----------------------------------------------------------------------

%Having established that the {\sc First} energy cutoff is effective in separating stronger from weaker hydrogen-bond constraints, albeit at a relatively low resolution, it seems sensible to step back and consider what may be said about the pattern of rigidity loss during dilution.
For the cytochromes that we have so far considered (\ref{sec-results-cytC}), the general pattern is one of gradual rigidity loss, particularly for $|E|>1$. This indicates a hierarchy of stability in the rigid clusters, with some areas being rigidified by very weak hydrogen bonds, some by bonds of medium strength and some by the strongest bonds. This is reminiscent of the gradual or second-order rigidity transition observed in some glassy networks \cite{SarWHT07}, specifically those with a wide diversity in their constraint distribution. Glassy networks with less diverse constraint networks, however, show a sudden, first-order-like rigidity transition in which the structure passes between largely flexible and largely rigid states on the addition or removal of only a few constraints.

In Figure \ref{fig-families} we show the patterns of rigidity loss for six different families of proteins as listed in Table \ref{Protein-Summary}. Our sample falls into two classes, those displaying a gradual pattern of rigidity loss (Figure \ref{fig-families}, (a) cytochrome C, (b) myoglobin, (c) lactalbumin, and (d) hemoglobin) and those displaying a sudden loss of rigidity once weak bonds are eliminated (Figure \ref{fig-families}, (e) HIV-1 protease and (f) trypsin). For proteins in this second class, all the $25$ structures that we examine display rapid loss of mainchain rigidity as weak bonds are removed and the mainchain has become almost entirely flexible once the cutoff energy is reduced below $-2$ kcal/mol. This indicates that the rigidity of clusters in these proteins is due to weaker hydrogen bonds and we do not see (as we do in the other four proteins) the persistence of rigid clusters bound by stronger hydrogen bonds.

The HIV-1 protease is a natural homodimer and we consider the rigidity of the dimer, as in \cite{JacRKT01}. For the other protein families our data is obtained from single protein chains; for example, for hemoglobin we analyse $\alpha$-hemoglobin chains. It should be clear that a protein chain treated in isolation always has fewer constraints then when treated as part of a complex, and indeed we find that the individual chains from the HIV-1 protease structure are even less rigid than the entire dimer (data in Supplementary Materials). For the case of hemoglobin, we can confirm that the rigidity of the isolated A-chain seen in Figure \ref{fig-families}(d) differs only slightly from the rigidity of the same chain when analysed as part of the full tetrameric hemoglobin structure (data in Supplementary Materials). Consideration of isolated HIV-1 protease monomers, or of full hemoglobin tetrameric complexes, thus does not alter their classification in terms of gradual or sudden loss of rigidity.

Comparison of these six protein families thus leads us to the conclusion that protein structures, like glassy networks, can display two distinct patterns of rigidity loss depending on the diversity of their constraint networks. We have identified two families of proteins, HIV protease and trypsin, whose members display rapid loss of rigidity as weaker hydrogen bonds are eliminated, in contrast to four other families of proteins which display a gradual loss of rigidity indicating a hierarchy of hydrogen-bond strengths in the constraints that maintain protein rigidity. 

% A third possible pattern would be a sudden loss of rigidity mediated by stronger hydrogen bonds, i.e.\ persistence of near-complete mainchain rigidity down to much lower cutoff values, but we do not observe this pattern among our sample.

%%%%%%%%%%%%%%%%%%%%%%%%%%%%%%%%%%%%%%%%%%%%%%%%%%%%%%%%%%%%%%%%%%%%%%%%
\section{Conclusion and outlook}
\label{sec-concl}
%%%%%%%%%%%%%%%%%%%%%%%%%%%%%%%%%%%%%%%%%%%%%%%%%%%%%%%%%%%%%%%%%%%%%%%%

Our motivation in this study was twofold: to clarify a methodological issue in the use of rigidity analysis on protein structures, by determining the robustness of RCDs against small structural variations and the significance of the cutoff energy value, and to obtain an insight into the patterns of rigidity loss during hydrogen-bond dilution, by comparison with the observed patterns in glassy networks.

On the first point, we find that there is considerable variation in the RCDs of structurally similar proteins during dilution. Figure \ref{fig-cyto-en}, for example, shows that among a group of cytochrome C structures drawn from similar eukaryotic mitochondria, energy cutoffs in the range from 0 to $-2$ kcal/mol (such as have typically been used for {\sc First}/{\sc Froda} simulations of flexible motion \cite{WelMHT05,JolWHTF06,JolWFT08,MacNBC07}) can produce a wide range of degrees of mainchain flexibility. We conclude that the results of rigidity analysis on individual crystal structures should not be over-interpreted as being ``the'' RCD for a protein. The hydrogen-bond energy function in {\sc First} is quite sensitive to small structural variations; while it successfully divides weaker from stronger bonds, it is not possible to identify a unique value of the hydrogen bond cutoff energy which can be applied to all protein structures to give meaningful results. Rather, each protein structure should first be subjected to rigidity dilution to produce a dilution plot; a suitable value of the cutoff energy can then be chosen to test a specific hypothesis about the rigidity and flexibility of the protein.
Similarly, when physical significance is attached to the pattern of rigidity loss \cite{HesRTK02}, then multiple similar examples of a given protein structure should be studied in order to be robust against structural variation.

% In combination with simulation of flexible motion, a ``hypothesis testing'' approach can reveal whether the presence or absence of a particular interaction is significant in allowing or forbidding a flexible motion to occur. An early use of this approach was the study by Jolley et al.\ \cite{JolWHTF06}, which compared two hypothetical pathways for formation of a protein complex by examining the changes in rigidity and flexibility which each pathway would entail.

On the second point, we find that proteins can display either gradual (second-order-like) or sudden (first-order-like) patterns of rigidity loss during dilution. We find sudden rigidity loss in two proteases, eukaryotic trypsin and viral HIV-1 protease.
Both consist largely of $\beta$-sheet secondary structure with little $\alpha$-helical content compared to the other proteins in our set, which may account for their different rigidity behaviour. Previous work \cite{RadHKT02} has emphasised the analogy between the rigidity transitions of proteins and of glassy networks; we have now found that the two distinct patterns of rigidity transition recently identified in glassy networks \cite{SarWHT07} are also seen in proteins.

%We have found on examining the HIV protease dimer structure that the dimer is more rigid at a given cutoff value than the isolated constituent monomers, due to the additional constraints arising from inter-chain interactions; however, this is only true for small values of the cutoff energy, consistent with previous studies on rigidity of complexes \cite{HesJT04}.

Our results in this paper suggest several avenues for further enquiry.
%A wider survey will allow us to develop a classification of proteins in terms of their pattern of rigidity loss and determine the significance of the different patterns.
The rigidity of protein monomers extracted from complexes should be systematically compared with their rigidity within the complex, which will be affected by interchain interactions. The robustness of flexible motion simulations based on rigidity analysis using different cutoff values must also be investigated. A recent study of the flexible motion of myosin \cite{SunRAJ08} found that the flexible motion of the myosin structure appeared qualitatively similar over a wide range of cutoff values covering both highly flexible and more rigid structures. This suggests that rigidity analysis retains its value as a natural coarse-graining for simulations even if the rigidity behaviour during dilution is as variable as we have found.

\ack
We thankfully acknowledge discussions with R.\ Freedman and T.\ Pinheiro. We  gratefully acknowledge financial support from the Leverhulme Trust (SAW and RAR, grant F/00 215/AH), BBSRC (JEJ) and EPSRC (JEJ and RAR, EP/C007042/1). We would like to thank two anonymous reviewers for their valuable comments and perspective.

%%%%%%%%%%%%%%%%%%%%%%%%%%%%%%%%%%%%%%%%%%%%%%%%%%%%%%%%%%%%%%%%%%%%%%%%
\appendix
%%%%%%%%%%%%%%%%%%%%%%%%%%%%%%%%%%%%%%%%%%%%%%%%%%%%%%%%%%%%%%%%%%%%%%%%
\section{Cutoff values in previous studies using {\sc First}}
\label{sec-previous}

Jacobs {\it et al.} \cite{JacRKT01}  comment that the results of {\sc First} analysis should  not be sensitive to the typical ``fluctuations known to occur within protein structures''. Their advice is that the cutoff should be at least $-0.1$ kcal/mol in order to eliminate a large number of very weak hydrogen bonds with energies in the range from $0.0$ to $-0.1$ kcal/mol, and that a natural choice is near the ``room temperature'' energy of $-0.6$ kcal/mol. As we have seen in section \ref{sec-vary}, this criterion is not sufficient to avoid sensitivity to small structural variations.

Rader {\it et al.} \cite{RadHKT02} consider the protein folding transition by monitoring $\langle r\rangle$ (mean number of bonded neighbours per atom) during rigidity dilution; they do not, however, comment on the hydrogen bond energy values. Hespenheide {\it et al.} \cite{HesRTK02} identify the protein folding core with ``the set of secondary structure that remain rigid the longest in the simulated denaturation'', without regard to the exact values of the cutoff energy at which rigidity is lost. Here the cutoff energy is used qualitatively to distinguish {\it weaker} from {\it stronger} bonds. In considering the rigidity of virus capsid protein complexes, Hespenheide {\it et al.} \cite{HesJT04} make use of a cutoff of $-0.35$ kcal/mol, a value chosen so that capsid protein dimers would be flexible while the inner ring of proteins in a pentamer of dimers would be rigid, and draw conclusions about the rigidity of other multimeric complexes. Meanwhile, Hemberg {\it et al.} \cite{HemYB06} use a different cutoff of $-0.7$ kcal/mol in a study on the dynamics of capsid assembly.

The {\sc Froda} geometric simulation algorithm \cite{WelMHT05} makes use of the RCD generated by {\sc First} as a coarse-graining. Simulations of protein mobility using {\sc First}/{\sc Froda} have tended to use cutoff values that are systematically lower than in applications of {\sc First} alone; typically $-1$ kcal/mol or lower \cite{WelMHT05,JolWHTF06,JolWFT08,MacNBC07,SunRAJ08}, as cutoff values closer to zero seem to include too many constraints to allow large-scale motion to occur. In a paper on the combination of rigidity analysis and elastic network modelling, Gohlke {\it et al.} \cite{GohT06} discuss RCDs of two protein crystal structures but do not specify a cutoff value, though the {\sc Froda} mobility simulations given in Figure 3a of that paper were performed using a cutoff of $-1.5$ kcal/mol and give an excellent match to experimental data from NMR ensembles.

%%%%%%%%%%%%%%%%%%%%%%%%%%%%%%%%%%%%%%%%%%%%%%%%%%%%%%%%%%%%%%%%%%%%%%%%
%\bibliographystyle{unsrt}\bibliography{bibliography/bibliograph}
%%%%%%%%%%%%%%%%%%%%%%%%%%%%%%%%%%%%%%%%%%%%%%%%%%%%%%%%%%%%%%%%%%%%%%%%

%\end{document}

\end{document}